\documentclass[12pt]{spieman}

\usepackage{amsmath,amsfonts,amssymb}
\usepackage{graphicx}
\usepackage{setspace}
\usepackage{tocloft}
\usepackage{lineno}
\usepackage{booktabs}
\usepackage{multirow}
\usepackage{enumitem}
\usepackage[table]{xcolor}
\usepackage{tabularx}
\usepackage{array}
\usepackage{hyperref}
\usepackage{float}
\usepackage{comment}
\usepackage{url}
\usepackage{longtable}

\title{ {From Flat-Optics Concept to Qualified Hardware:} Skills Map for the Meta-Optics and Diffractive Optics Workforce}

\author[a]{Ingrid Torres}
\author[a,*]{Alex Krasnok}
\affil[a]{Florida International University, Electrical \& Computer Engineering, Miami, Florida, USA}

\cftpagenumbersoff{figure}
\cftpagenumbersoff{table}

\newcommand{\Lone}{\textbf{L1}}
\newcommand{\Ltwo}{\textbf{L2}}
\newcommand{\Lthree}{\textbf{L3}}

\definecolor{p1}{gray}{0.90}
\definecolor{p2}{gray}{0.80}
\definecolor{p3}{gray}{0.65}

\newcommand{\placeholderbox}[2]{%
\fbox{\parbox[c][#1][c]{0.95\linewidth}{\centering #2}}}

\begin{document}
\doublespacing
\maketitle

\begin{abstract}
Flat optics is now judged by more than a strong simulation or a single laboratory demonstration. To reach release, a device must survive a chain of handoffs: requirements, model selection, verification, layout release, fabrication, calibrated validation, packaging, and qualification. Diffractive optics brings mature routes for beam shaping and compact wavefront control, while meta-optics expands the design space through wavelength-scale control of phase, amplitude, and polarization. In both families, projects often slow down not because the optical function is impossible, but because the evidence required at each handoff is incomplete, poorly documented, or mismatched to the next decision.  {This tutorial organizes that problem into a stage-gate workflow, a set of compact technical checks, worked device examples, an artifact-based skills map, and an educational translation into workforce models, course deliverables, and assessment logic. The emphasis is practical: reduce avoidable redesign loops, make performance claims auditable, and clarify what students, instructors, and employers should be able to produce, review, and approve. The broader aim is to make the path from flat-optics concept to qualified hardware easier to understand, easier to teach, and easier to repeat.}
\end{abstract}


\section{\large Introduction}

Flat optics no longer faces only a physics question. The harder question is whether a thin patterned surface can move from concept to released hardware without losing optical intent along the way. A promising design now has to survive requirement flow-down, model choice, mask release, fabrication, calibrated validation, packaging, and qualification. That challenge has become central as cameras, sensors, displays, spectrometers, beam-steering systems, and wearable optics all demand more function in less space \cite{Kildishev2013PlanarPhotonics,Kuznetsov2024RoadmapACSP,Schulz2024RoadmapAPL,Yang2025CommercializingMetasurfaces}. Diffractive optics already appears at scale in structured-light projectors for consumer depth sensing, and diffractive waveguide combiners are established in commercial augmented-reality platforms \cite{LightTrans2019DotProjector,Tian2025ARWaveguideLSA}. Meta-optics is now moving through the same transition from laboratory result to product candidate.

That transition has two historical roots. Diffractive optics grew from holography and interference-based wavefront shaping, from Gabor's foundational idea through off-axis reconstruction, computer-generated holograms, and multilevel diffractive optical elements that became useful in testing, beam shaping, displays, and spectroscopy \cite{Gabor1948MicroscopicPrinciple,LeithUpatnieks1962Reconstructed,LeithUpatnieks1964Diffused,LohmannParis1967Binary,Zhang2023DOE75Years}. Meta-optics emerged from metamaterials and accelerated when metasurfaces showed that a surface much thinner than a wavelength could impose strong, designed phase changes across an aperture \cite{Pendry2000PerfectLens,Smith2000LeftHanded,Yu2011PhaseDiscontinuities,Yu2014FlatOptics,Chen2016MetasurfaceReview}. Dielectric gradient metasurfaces, high-contrast transmitarrays, and polarization-aware dielectric metasurfaces then showed that flat optics could combine strong field control with high transmission and fabrication routes that are increasingly realistic \cite{Lin2014DielectricGradient,Arbabi2015Transmitarrays,Arbabi2015CompleteControl,Decker2015Huygens}. Visible-wavelength metalenses and later achromatic metalenses pushed the story further by showing that flat optics could address problems once reserved for much thicker optical trains \cite{Khorasaninejad2016VisibleMetalenses,Chen2018BroadbandAchromatic,Wang2018BroadbandAchromatic,Shrestha2018BroadbandAchromatic,Balli2020HybridAchromatic}.

The two families are best treated as complementary engineering options rather than as simple competitors. Diffractive optics remains especially attractive when Fourier-optics design, large apertures, phase quantization, and scalable replication dominate the problem. Meta-optics becomes especially valuable when phase, amplitude, and polarization must be controlled together, or when the desired function depends on wavelength-scale scattering that a simple relief profile cannot easily supply \cite{Yu2014FlatOptics,Chen2016MetasurfaceReview,Kuznetsov2024RoadmapACSP}. In practice, the boundary is set less by labels than by fabrication limits, metrology burden, packaging, yield, and cost. Metagratings and hybrid architectures make that overlap explicit: many useful devices already borrow the language of both families \cite{Zhang2020MetagratingCPD,Arbabi2015Transmitarrays,Balli2020HybridAchromatic}. Figure~\ref{fig:history_timeline} summarizes that historical arc on one page.
\begin{figure}[t]
\centering
\IfFileExists{Fig1.png}{
\includegraphics[width=1\linewidth]{Fig1.png}
}{
\placeholderbox{2.2in}{Historical development placeholder}
}
\caption{Historical development of diffractive optics and meta-optics (1948--2026). Lane A tracks holography and diffractive optics toward mature DOE-based industrial components. Lane B tracks metamaterials and metasurfaces toward metalenses and multifunctional meta-systems. The shaded deployment era highlights the shift from demonstrations to scalable fabrication, manufacturable design, calibrated metrology with uncertainty, packaging, reliability, and commercialization.}
\label{fig:history_timeline}
\end{figure}

The central question has changed with the field. Early papers often asked whether a flat optic could perform a target function at all. The more important question now is whether that function can be repeated, measured, packaged, and qualified within a real product cycle. Recent roadmaps and commercialization-focused reviews keep returning to the same obstacles: scalable fabrication, design for manufacturing, application-specific metrology, packaging, reliability, and system integration \cite{Kuznetsov2024RoadmapACSP,Schulz2024RoadmapAPL,Yang2025CommercializingMetasurfaces,Kossowski2025MetrologyNPJ,Dirdal2020HighThroughputMetalens,Guo2007NILReview}. The field is no longer asking only whether a device works once. It is asking whether the full path to release is defensible.

 {This paper addresses that release problem. It is not a broad survey of all flat-optics physics. Instead, it follows the more decisive story: how an optical idea becomes a part that another team can trust, fabricate, measure, package, and release. The paper is organized in three connected parts. Sections~2--4 follow the device journey from requirements to validation and packaging. Sections~5--6 define the skills and role targets needed to keep those handoffs reliable. Sections~7--10 translate that logic into curriculum design, training, and implementation. That structure is meant to help three audiences at once: hiring teams that need sharper role language, instructors who need a bridge from theory to handoff-ready artifacts, and students or practicing engineers who need a more honest way to judge whether their work can survive transfer to the next stage.}

Despite their structural differences, meta-optics and diffractive optics can still be written in a common optical language. A thin optical element is often described by its complex transmission function,
\begin{equation}
t(x,y)=A(x,y)\exp\!\left[i\phi(x,y)\right],
\label{eq:complex_transmission_intro}
\end{equation}
where \(A(x,y)\) is the transmitted amplitude and \(\phi(x,y)\) is the transmitted phase at position \((x,y)\). A focusing element aims to impose a lens phase profile such as
\begin{equation}
\phi_{\mathrm{lens}}(x,y)=-\frac{2\pi}{\lambda_0}\left(\sqrt{x^2+y^2+f^2}-f\right),
\label{eq:lens_phase_intro}
\end{equation}
where \(f\) is the focal length.  {Unless otherwise stated, \(\lambda_0\) denotes the wavelength in vacuum and \(k_0=2\pi/\lambda_0\). For propagation in a homogeneous medium of refractive index \(n_p\), the wavelength in that medium is \(\lambda_m=\lambda_0/n_p\). We use compact indices throughout: \(n_d\) for the device material and \(n_0\) for the surrounding medium. When the propagation medium is the same as the surrounding medium, \(n_p=n_0\).} A classical diffractive lens, a multilevel DOE, and a metalens may all target Eq.~(\ref{eq:lens_phase_intro}); what changes is how the phase is produced, which approximations remain honest, and how sensitive the final structure becomes to fabrication and packaging \cite{GoodmanFourierOptics,BornWolf,Yu2014FlatOptics,Chen2016MetasurfaceReview,Zhang2023DOE75Years}.

In meta-optics, that target field is realized with arrays of subwavelength scatterers. The phase may come from propagation through a high-index element, from a resonance, or from in-plane rotation. For circularly polarized light, the Pancharatnam--Berry phase is
\begin{equation}
\phi_{\mathrm{PB}}=\pm 2\theta,
\label{eq:pb_phase_intro}
\end{equation}
where \(\theta\) is the in-plane rotation angle. Because metasurfaces often control phase, amplitude, and polarization together, they usually require vector electromagnetic models rather than scalar phase-only masks \cite{Yu2011PhaseDiscontinuities,Yu2014FlatOptics,Chen2016MetasurfaceReview,Genevet2017PlanarOptics,Arbabi2015CompleteControl}.

In diffractive optics, the target field is more often realized through an engineered phase or amplitude profile. In a transmissive relief DOE, a local height \(h\) produces the phase delay
\begin{equation}
 {\phi(h)=\frac{2\pi}{\lambda_0}\left(n_d-n_0\right)h,}
\label{eq:height_phase_intro}
\end{equation}
where \(n_d\) is the refractive index of the DOE material and \(n_0\) is the refractive index of the surrounding medium. For a DOE operating in air, \(n_0\approx 1\). A grating with period \(\Lambda\) follows
\begin{equation}
 {m\lambda_0=n_0\Lambda\left(\sin\theta_m-\sin\theta_i\right),}
\label{eq:grating_intro}
\end{equation}
 {where the signs of \(\theta_i\) and \(\theta_m\) are defined by the tangential component of the wave vector. Other common angle conventions lead to algebraically equivalent plus/minus forms. Many DOEs can be designed first with scalar Fourier optics.} That route is fast and often adequate when polarization conversion is not part of the requirement and the feature size is not deeply subwavelength. If fabrication replaces a continuous phase profile by \(N\) discrete height levels, the ideal first-order diffraction efficiency is approximately
\begin{equation}
\eta_N\approx\left[\frac{\sin(\pi/N)}{\pi/N}\right]^2,
\label{eq:quantization_efficiency_intro}
\end{equation}
which gives a practical trend: more phase levels can raise efficiency, but they also increase fabrication complexity and sensitivity to profile error \cite{GoodmanFourierOptics,BornWolf,Zhang2023DOE75Years}.

\begin{figure}[t]
\centering
\IfFileExists{Fig2.png}{
\includegraphics[width=0.9\linewidth]{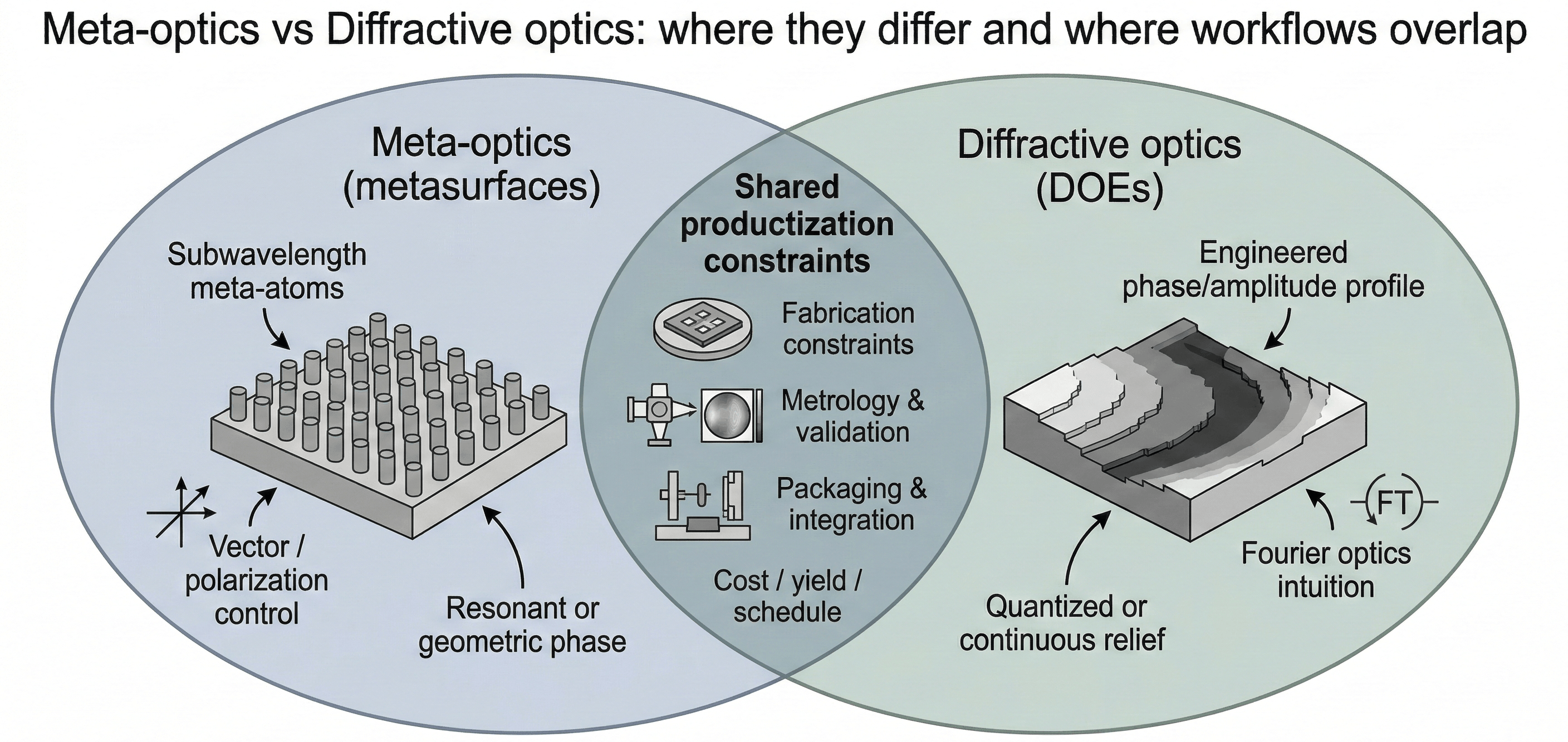}
}{
\placeholderbox{2.1in}{Meta-optics and diffractive optics placeholder}
}
\caption{Meta-optics and diffractive optics pursue the same engineering goal---shaping the optical field across a surface---but do so with different physical building blocks. Metasurfaces rely on wavelength-scale scatterers and naturally bring amplitude and polarization into the design space. Diffractive optics relies on engineered phase or amplitude profiles and often begins with Fourier-optics design intuition. In products, the two overlap because both must eventually satisfy the same constraints on fabrication, metrology, packaging, yield, cost, and schedule.}
\label{fig:VnnDgm}
\end{figure}

The overlap between the two families matters more than the label. A flat lens or beam deflector may target the same focal length, aperture, field of view, efficiency, or steering angle while relying on very different physical structures underneath. The numerical aperture is
\[
\mathrm{NA}=n_0\sin\theta_{\max},
\]
where \(n_0\) is the refractive index of the surrounding medium. As numerical aperture rises, bandwidth broadens, or polarization becomes part of the function, the need for vector modeling and careful validation usually increases \cite{Chen2016MetasurfaceReview,Kossowski2025MetrologyNPJ,Zhang2023DOE75Years}. Figure~\ref{fig:VnnDgm} previews the core point of this tutorial: even when the optical concept is sound, projects lose time when the workflow between modeling, release, process control, metrology, and packaging breaks down.

With that common language in place, the next section follows the stage-gate path that turns a flat-optics idea into a part another team can fabricate, test, and trust.

\section{\large From requirements to release: productization workflow}

\begin{figure}[t]
\centering
\IfFileExists{Figure3.png}{
\includegraphics[width=1\linewidth]{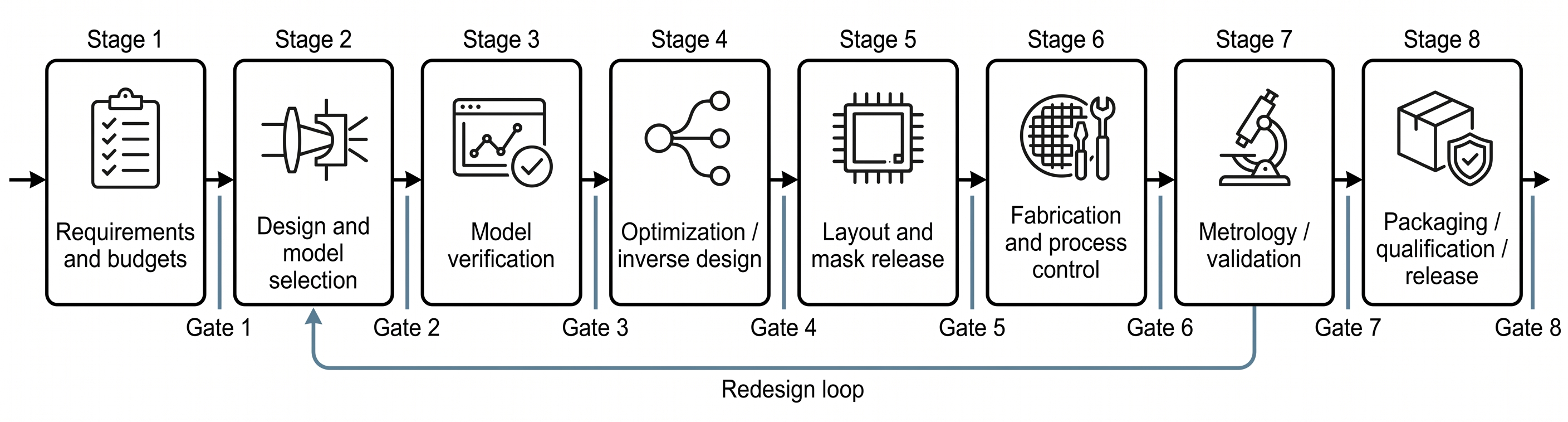}
}{
\placeholderbox{2.25in}{Workflow figure placeholder}
}
\caption{ {High-level stage-gate workflow for meta-optics and diffractive optics. The figure shows the stage order, the redesign loop, and the position of Gates~1--8. Gate~7 sits between metrology/validation and packaging/qualification/release. Detailed artifacts and exit questions are listed in Table~\ref{tab:stagegates}.}}
\label{fig:PdzWrkflw}
\end{figure}

Every released flat optic begins with a promise and ends with a documented decision. Between those two points, the promise has to be translated into budgets, models, layouts, process controls, measurements, and qualification evidence. Figure~\ref{fig:PdzWrkflw} shows that journey as a stage-gate workflow. It makes the sequence visible, identifies the redesign loop, and places Gates~1--8 in the order they are typically encountered. Table~\ref{tab:stagegates} then adds the minimum artifact and exit question at each gate.

In this tutorial, productization means the sequence of decisions that turns a desired optical function into a part that can be fabricated, measured, integrated, and released. The sequence starts with requirement flow-down. A system-level need such as field of view, focusing efficiency, diffraction angle, spectral range, image quality, or allowable thermal drift must be translated into device-level quantities such as focal length, numerical aperture, phase error, polarization leakage, minimum feature size, alignment tolerance, and pass/fail limits. If that translation is vague at the start, surprises usually appear at the end.

A useful discipline is to make each requirement trace in three directions. It should trace downward to a device quantity that can be designed, sideways to a process or packaging variable that can disturb it, and forward to a measurement that can validate it. A traceability matrix is simply the place where those links are written down. For a steering requirement, for example, the matrix should connect the target angle to a phase slope or grating period, to the fabrication tolerance that can shift that geometry, and to an angle-resolved measurement plan with a stated reference and acceptance limit. The same logic applies whether the device is a structured-light DOE, a metalens, or a polarization-selective metasurface.

Verification and validation belong to different parts of the story. Verification asks whether an artifact meets the requirement written for it. Validation asks whether the resulting hardware is fit for its intended use in the stated context \cite{ISO24765Vocabulary2017,JCGM100_2008,JCGM106_2012}. In flat optics, verification governs model credibility, release correctness, and process discipline. Validation governs hardware performance claims. Because validation decisions depend on uncertainty and risk, this paper treats uncertainty budgets and explicit decision rules as required evidence at Gate~7, not as optional paperwork added later.

The redesign loop is equally important. Measured disagreement does not automatically mean the original optical concept was wrong. The gap may have entered through the model, the release package, the process window, the test definition, or the package itself. A useful workflow does more than say ``iterate.'' It helps the team locate the source of the gap and return to the correct stage with a written reason. Once the gates are explicit, the next question is which compact technical checks actually change decisions at those gates.

\section{\large Technical checks that change gate decisions}

At each gate, a small number of questions determine whether the project moves forward, loops back, or accumulates hidden risk. Those questions sit between the optical physics and the product constraints shown in Fig.~\ref{fig:VnnDgm}: fabrication limits, metrology burden, packaging, yield, cost, and schedule. The detailed workflow checklist remains in the Supplementary Material. Here the focus stays on the compact checks that repeatedly influence real decisions.

These relations do not replace full-wave simulation or careful measurement. Their value is different. They force the team to state what quantity is being specified, which regime is being assumed, and whether the reported result is actually the one the requirement asked for. That discipline is often what prevents late-stage confusion.

The first screening question is simple: what does the reported spot size mean? For a circular pupil, the radius of the Airy disk out to the first zero is approximately
\begin{equation}
r_{\mathrm{Airy}}\approx 0.61\,\frac{\lambda_0}{\mathrm{NA}},
\label{eq:airy_radius}
\end{equation}
and the intensity full width at half maximum diameter is approximately
\begin{equation}
\mathrm{FWHM}_{\mathrm{diam}}\approx 1.03\,\frac{\lambda_0}{\mathrm{NA}}.
\label{eq:fwhm_diameter}
\end{equation}
These two numbers are easy to blur together, especially when spot size, resolution, and collection geometry are discussed loosely. At a gate review, they matter less as textbook derivations than as a way to force agreement on the exact performance quantity being claimed.

\begin{figure}[t]
\centering
\IfFileExists{Figure4.png}{
\includegraphics[width=0.82\linewidth]{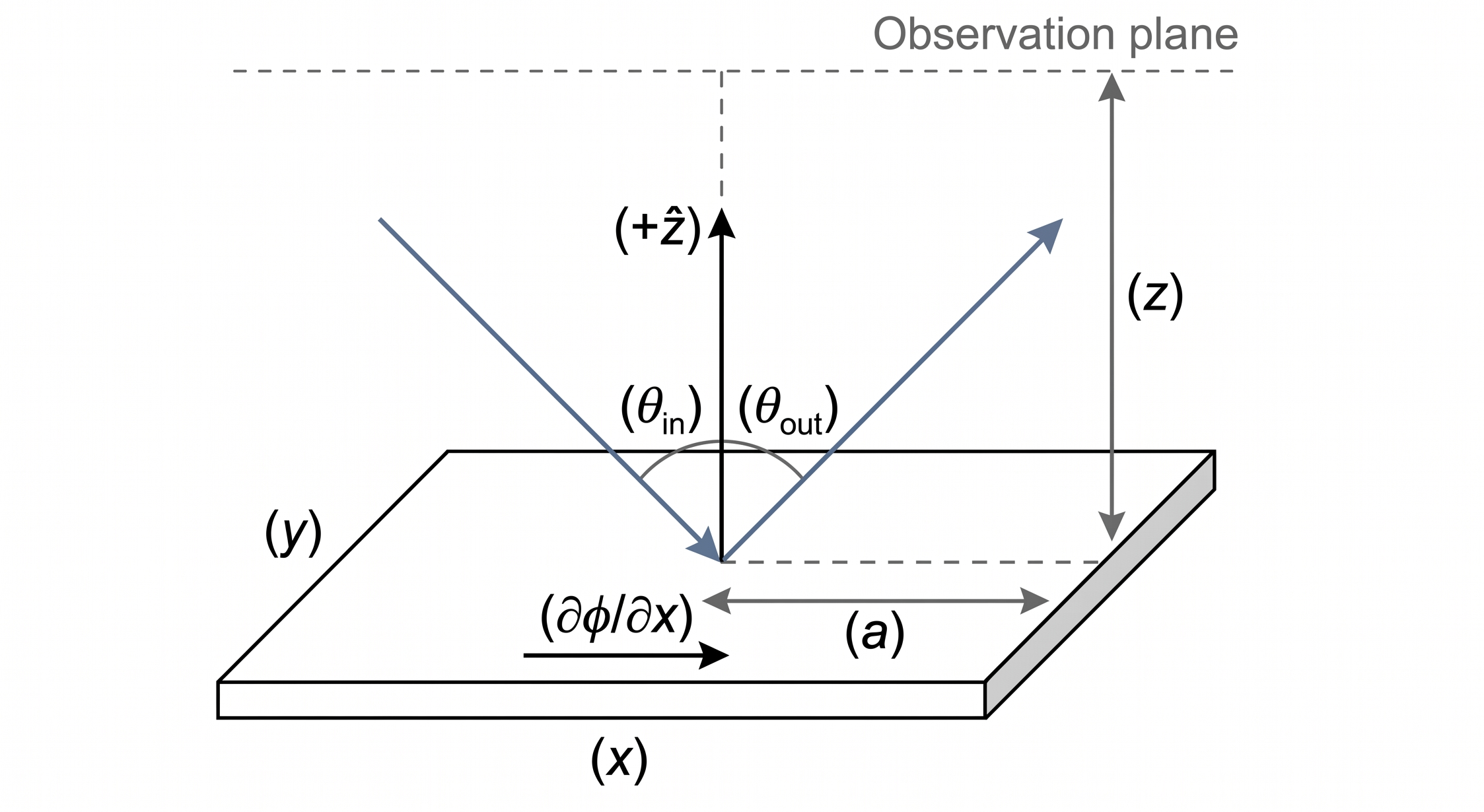}
}{
\placeholderbox{2.0in}{Geometry figure placeholder}
}
\caption{ {Geometry used in Eqs.~(\ref{eq:fresnel_number})--(\ref{eq:phase_gradient_angle}). The flat optic lies in the \(x\)-\(y\) plane, the surface normal is \(+\hat{z}\), \(z\) is the propagation distance measured along that normal, \(a\) is the half-aperture used in the Fresnel number, and \(\theta_{\mathrm{in}}\) and \(\theta_{\mathrm{out}}\) are measured from the surface normal. The schematic defines a generic surface-based angle convention; the same notation supports transmission or reflection once the corresponding sign convention is chosen.}}
\label{fig:phase_gradient_geometry}
\end{figure}

The next question is geometric: is the chosen propagation picture even honest? Figure~\ref{fig:phase_gradient_geometry} defines the surface-based convention used for the propagation and phase-gradient relations below.

The Fresnel number is
\begin{equation}
 {F=\frac{a^2}{\lambda_m z},}
\label{eq:fresnel_number}
\end{equation}
 {where \(a\) is a characteristic half-aperture, \(z\) is the propagation distance measured along \(+\hat{z}\), and \(\lambda_m=\lambda_0/n_p\) is the wavelength in the propagation medium of refractive index \(n_p\). When \(F\ll 1\), a far-field picture is often sufficient. When \(F\gtrsim 1\), near-field evolution matters and the metrology geometry should be chosen accordingly. This is an early screening check, not a substitute for a full validation plan.}

A third question concerns steering or beam deflection: is the requested angle compatible with a realizable phase gradient? Tangential momentum conservation gives
\begin{equation}
k_{x,\mathrm{out}}=k_{x,\mathrm{in}}+\frac{\partial \phi}{\partial x},
\label{eq:phase_gradient_kx}
\end{equation}
and, for a transmissive device in equal surrounding media on both sides of the surface,
\begin{equation}
 {\sin\theta_{\mathrm{out}}-\sin\theta_{\mathrm{in}}=
\frac{\lambda_0}{2\pi n_0}\,\frac{\partial\phi}{\partial x}.}
\label{eq:phase_gradient_angle}
\end{equation}
 {Equation~(\ref{eq:phase_gradient_angle}) is written for one common transmission convention. Reflective devices and alternate sign conventions lead to algebraically related forms. These are screening relations. They tie steering angle, phase slope, period, and feature size together in one line, and they expose impossible or high-risk targets before mask release.}

Before any team trusts an efficiency number, it helps to write down what is being counted.  {In this paper, \(R\), \(T\), and \(A\) denote the total reflected, transmitted, and absorbed power fractions, respectively, each normalized to the incident power. When diffraction is present, \(R\) and \(T\) include the relevant orders. The basic conservation check is}
\begin{equation}
R+T+A=1,
\label{eq:energy_balance}
\end{equation}
and it often reveals missing orders in simulation, incomplete normalization in measurement, or stray collection loss.

Manufacturability also needs to enter the story early.  {Minimum width, spacing, aspect ratio, bias, and loading limits should appear at Gate~1 as part of the requirement envelope and at Gate~2 as part of model and parameterization choice. Gate~4 is where those limits are enforced inside the optimization and checked for robustness. If the constrained design drifts far from the original baseline, the team should return to Gates~1--2, update the traceability matrix and budgets, and re-baseline before release. That point matters especially in inverse design, where late-added manufacturability filters can qualitatively change the solution \cite{Schubert2022StrictFoundry,Augenstein2020StructuralIntegrity,Tanriover2023ManufacturableMetasurfaces}.}

A first estimate of phase sensitivity to fabrication bias is
\begin{equation}
\Delta\phi\approx
\frac{\partial\phi}{\partial w}\,\Delta w+
\frac{\partial\phi}{\partial h}\,\Delta h,
\label{eq:phase_sensitivity}
\end{equation}
where \(w\) and \(h\) represent a feature width and height. This relation does not solve the process problem by itself. Its value is practical: it connects geometry drift to optical risk in a form that supports robustness studies and process-window discussions.

Once hardware is measured, the workflow has to turn numbers into decisions. For a derived quantity \(y\), a first-order estimate of the standard uncertainty is
\begin{equation}
u_y^2\approx \sum_i\left(\frac{\partial y}{\partial x_i}\right)^2u_{x_i}^2,
\label{eq:uncertainty_standard}
\end{equation}
and a one-sided acceptance rule with guard band is
\begin{equation}
y_m+k\,u_y\le y_{\max},
\label{eq:decision_rule}
\end{equation}
where \(y_m\) is the measured value, \(u_y\) is the standard uncertainty, \(k\) is a chosen coverage factor, and \(y_{\max}\) is the upper specification limit \cite{JCGM100_2008,JCGM106_2012}. The point is not to perform uncertainty math for its own sake. The point is to make Gate~7 auditable.

 {A claim becomes useful only when it is tied to a measurement plan. Table~\ref{tab:claimmetrology} maps common flat-optics claims to the measurement logic needed to defend them. The goal is not to replace detailed test plans. The goal is to make clear that a reported number is inseparable from its reference, dominant uncertainty, and gate context \cite{Kossowski2025MetrologyNPJ,JCGM100_2008,JCGM106_2012}.}

\begin{table}[t]
\caption{ {Claim-to-metrology map for common flat-optics performance statements. The table links each claim to a suitable primary method, the reference or normalization that must be stated, the dominant uncertainty contributors, and the gate where the evidence usually matters most.}}
\label{tab:claimmetrology}
\centering
\footnotesize
\renewcommand{\arraystretch}{1.08}
\setlength{\tabcolsep}{4pt}
\begin{tabularx}{\textwidth}{>{\raggedright\arraybackslash}p{2.8cm} >{\raggedright\arraybackslash}p{3.0cm} >{\raggedright\arraybackslash}p{3.1cm} >{\raggedright\arraybackslash}X >{\raggedright\arraybackslash}p{1.2cm}}
\toprule
\textbf{Claim} & \textbf{Typical primary method} & \textbf{Reference or normalization} & \textbf{Dominant uncertainty contributors} & \textbf{Gate}\\
\midrule
Diffraction or steering angle & Fourier-plane imaging or goniometric measurement & Angular calibration, pixel-to-angle map, optical-axis definition & Alignment, working distance, centroiding, lens distortion & G7\\
Focusing efficiency & Power in the stated focal region or collection NA & Incident power, background subtraction, collection geometry, spectral condition & Source stability, detector calibration, aperture definition, stray light & G7\\
Wavefront error or focal shift & Interferometry, Shack--Hartmann sensing, or through-focus scan & Reference flat or reference optic, environmental baseline & Calibration drift, fit model, stage repeatability, temperature & G7/G8\\
Polarization leakage or conversion & Analyzer sweep, Stokes/Jones measurement & Input-state calibration, analyzer extinction, basis definition & Polarization purity, extinction ratio, alignment, retardance drift & G7\\
Spectral response or bandwidth & Spectrally resolved transmission, reflection, or functional scan & Source spectrum, detector response, wavelength calibration & Spectral calibration, normalization, coherence effects & G7\\
Package-induced drift & Repeated optical test under thermal, mechanical, or humidity stress & Pre-package baseline, fixture reference, stress history & Fixture repeatability, thermal lag, adhesive creep, relaxation history & G8\\
\bottomrule
\end{tabularx}
\end{table}

Table~\ref{tab:stagegates} turns that logic into the minimum evidence a reviewer should be able to inspect before the next team proceeds.
\begin{table}[t]
\caption{High-level gate summary aligned to Fig.~\ref{fig:PdzWrkflw}. Figure~\ref{fig:PdzWrkflw} shows the flow only. This table lists the minimum artifact and exit question at each gate. The detailed checklist remains in SM Table~S1.}
\label{tab:stagegates}
\centering
\footnotesize
\renewcommand{\arraystretch}{1.06}
\setlength{\tabcolsep}{4pt}
\begin{tabularx}{\textwidth}{
>{\raggedright\arraybackslash}p{0.9cm}
>{\raggedright\arraybackslash}p{2.3cm}
X
>{\raggedright\arraybackslash}p{5.8cm}}
\toprule
\textbf{Gate} & \textbf{Stage} & \textbf{Minimum artifact} & \textbf{Exit question / primary reviewer}\\
\midrule
G1 & Requirements \& budgets &
Requirements spec, optical budget, traceability matrix, initial fabrication envelope &
Are the requirements measurable, traceable, and consistent with the available process and test envelope? / Systems or optical architect \\

G2 & Design \& model selection &
Assumptions log, model choice, scalar/vector and periodic/finite justification, material-data sources, process assumptions &
Does the selected physical picture match the function, geometry scale, polarization behavior, and manufacturing limits? / Design + simulation lead \\

G3 & Model verification &
Convergence evidence, energy-balance check, independent cross-checks, verified model note &
Can the model be trusted for release-driving decisions? / Design + simulation lead \\

G4 & Optimization / inverse design &
Formal constraint set, reproducible optimization archive, robustness study, final independent verification &
Are fabrication constraints enforced, and does the constrained design still meet the requirement? If not, return to G1--G2 and re-baseline. / Design + simulation lead \\

G5 & Layout \& mask release &
GDS/OASIS, layer map, release README, DRC report, revision tag &
Could another team fabricate this without clarification? / Process or nanofab lead \\

G6 & Fabrication \& process control &
Process traveler, process window, inline metrology plan, deviation handling note &
Are critical process controls linked to optical impact? / Process or nanofab lead \\

G7 & Metrology / validation &
Calibrated dataset, uncertainty budget, decision rule, pass/fail report &
Does measured hardware meet the requirement within stated uncertainty? / Metrology lead \\

G8 & Packaging / qualification / release &
Alignment/drift budget, stress-to-metric map, qualification report, release record &
Does packaged hardware remain fit for intended use across expected stresses? / Manufacturing, quality, or program owner \\
\bottomrule
\end{tabularx}
\end{table}

With that structure in place, the modeling choices become easier to place. Scalar Fourier-optics models often serve classical gratings and DOEs well when polarization conversion is not part of the requirement. Vector models are required when the function itself depends on polarization mixing, anisotropy, or resonant scattering. The relevant question at Gate~2 is not which solver is fashionable. It is which physics must be explicit to defend the claim.

RCWA and related Fourier-modal methods are effective where periodicity and layered structure dominate. Their practical value often comes from building unit-cell libraries that map geometry to phase, amplitude, and polarization response. But the release-driving artifact is not a single RCWA plot. It is a verified periodic-model package with convergence evidence, energy balance, and one clear statement of what periodic boundary conditions do and do not represent for the final device \cite{Lalanne1996RCWA,Li1996FourierSeries,Moharam1995RCWA,LiuFan2012S4,HugoninLalanne2021Reticolo}.

Full-wave solvers such as FDTD and FEM become necessary when edge effects, broad bandwidth, cover layers, adhesives, housing boundaries, or material loss shift the response in a meaningful way. The relevant artifact is a verified model package with mesh studies, boundary-condition justification, and at least one rerunnable regression case \cite{Oskooi2010Meep}. In practice, that discipline matters as much as the solver name.

Layout release, fabrication, metrology, and packaging should all be treated as optical steps rather than as downstream administration. A release package must be readable without oral rescue. A process window must map geometry drift to optical drift. A validation report must state the measured quantity, reference, calibration, uncertainty, and decision rule. A qualification plan must map each applied stress to an optical metric. These are the habits that move a device from an attractive design to a reproducible product \cite{KLayoutDRCManual,GDSFactoryDocs,Dirdal2020HighThroughputMetalens,Guo2007NILReview,Schulz2024RoadmapAPL,Yang2025CommercializingMetasurfaces}. Those checks become easier to trust in context, so the next section follows three short device stories that show where projects usually fail and how the redesign loop should respond.

\section{\large Worked examples across the workflow}

 {A framework becomes useful when it is placed under load. Table~\ref{tab:workedexamples} distills three common flat-optics stories into one page. The aim is not to report three full product programs. The aim is to show how the same gate logic behaves differently for a scalar-first DOE, a vector-critical metasurface, and a device that passes bench validation but fails after packaging \cite{Song2019DOEStructuredLight,Hsu2024MetasurfaceStructuredLight,Balli2020HybridAchromatic,Schubert2022StrictFoundry}.}

\begin{table}[t]
\caption{ {Three concise workflow examples showing how the same gate logic behaves for different device classes. The examples are written to show which artifacts become decisive, where failures usually enter, and how the redesign loop is triggered.}}
\label{tab:workedexamples}
\centering
\footnotesize
\renewcommand{\arraystretch}{1.08}
\setlength{\tabcolsep}{4pt}
\begin{tabularx}{\textwidth}{>{\raggedright\arraybackslash}p{2.7cm} >{\raggedright\arraybackslash}p{3.3cm} X >{\raggedright\arraybackslash}p{3.2cm}}
\toprule
\textbf{Case} & \textbf{What drives the early design} & \textbf{Artifact that becomes decisive} & \textbf{Typical redesign trigger}\\
\midrule
DOE structured-light projector & A field of view, dot density, and source wavelength that can often be screened with scalar phase-gradient and diffraction relations & A traceability matrix linking pattern geometry, fabrication bias, source conditions, and angle-resolved validation; then a release package that preserves phase-level intent & The far-field pattern matches the nominal angle but fails dot uniformity, order balance, or encoded-pattern decoding after fabrication \\

Polarization-selective metasurface or metalens & A function that cannot be stated honestly without vector response, polarization basis, and a finite-aperture credibility check & A verified model package that includes convergence, one independent cross-check, polarization-aware outputs, and a test plan that measures the same basis used in design & Finite-aperture behavior or cross-polar leakage departs from the unit-cell library, even though the nominal phase response looked correct \\

Packaged flat optic & A bench-top optic that initially meets focal, angular, or wavefront targets before assembly and environmental stress & An alignment budget and a stress-to-metric map that translate cover layers, adhesive, tilt, and temperature into optical drift & Gate~7 passes on unpackaged hardware, but Gate~8 fails because packaging shifts focal position, polarization response, or steering angle beyond budget \\
\bottomrule
\end{tabularx}
\end{table}

The DOE example is the clearest place to start. A structured-light projector or beam deflector can often be screened with scalar diffraction relations before more detailed simulation begins. That simplicity is useful, but it can hide the real risk. The device may meet the nominal steering angle and still fail after fabrication because dot uniformity, order balance, or coded-pattern decoding falls outside the usable range. In practice, the decisive artifact is rarely the scalar model alone. It is the traceability matrix that links pattern geometry, source conditions, fabrication tolerance, and the exact validation geometry used to accept the part \cite{Song2019DOEStructuredLight}.

The metasurface example reveals a different failure mode. A polarization-selective beam splitter, waveplate-like element, or polarization-aware metalens cannot be described honestly with a scalar phase mask. The early risk is not only whether the unit cell delivers the right phase. It is whether the finite device, the chosen polarization basis, and the measurement definition still refer to the same object. That is why the decisive artifact is a verified vector model package with convergence, at least one independent cross-check, and a test plan that preserves the analyzer and input-state definitions used in design \cite{Hsu2024MetasurfaceStructuredLight,Kossowski2025MetrologyNPJ}.

The packaging example is often the most instructive because it surprises new teams. A flat optic may pass its bench-top validation and still fail at release because a cover layer, adhesive, tilt, or mechanical stress shifts the effective response. In that situation, the redesign loop should not default to ``change the pattern.'' It should first ask whether the failing quantity belongs to the design, the package, or the test context. The decisive artifact is therefore not only the Gate~7 validation report. It is the alignment budget and stress-to-metric map that explain why a device that worked as hardware did not survive becoming a product \cite{Balli2020HybridAchromatic,Yang2025CommercializingMetasurfaces}.

These examples identify the artifacts that matter most at failure points. The next question is who on the team must be able to create, review, and approve those artifacts.

\section{\large How the skills map is constructed and kept current}

 {This section presents a tutorial framework grounded in literature and productization practice, not a labor-market survey or a Delphi-style consensus study. The skills map is built from four evidence streams: recent technical roadmaps and synthesis reviews, foundational optics and numerical-method literature, standards and workflow documents, and workforce or competency frameworks \cite{Kuznetsov2024RoadmapACSP,Schulz2024RoadmapAPL,Zhang2023DOE75Years,Kossowski2025MetrologyNPJ,GoodmanFourierOptics,BornWolf,ISO24765Vocabulary2017,JCGM100_2008,JCGM106_2012,AIMWorkforceRoadmap,ONETContentModel,SFIA9}. A source is included only if it helps answer at least one practical question: which stage is under discussion, which decision is made there, which artifact must exist, which failure mode is common, or what evidence separates guided work from independent work or sign-off work.}

The key translation step is to rewrite source material in artifact language rather than topic language. A solver paper is useful here only if it clarifies what a credible model package must contain. A metrology paper is useful only if it clarifies what a decision-ready validation report must contain. That coding step turns literature into a reusable skills map instead of a reading list.

The proficiency levels are defined once here and used consistently throughout the paper. \Lone\ means the person can contribute under guidance. \Ltwo\ means the person can deliver a complete, handoff-ready artifact that another role can use without oral rescue. \Lthree\ means the person can review, approve, and sign off. These definitions matter because the paper is not trying to rank people by topic familiarity; it is trying to clarify what can safely cross a gate review.

A practical map also needs a practical refresh cycle. Tool examples will change. Acceptance checks will evolve as standards and qualification expectations shift. Different sectors will place different weight on packaging, manufacturing, or metrology over time. The stable part of the framework is the gate logic. The changing part is how examples, tools, and expectations are versioned around that logic. With those definitions in place, the next section can be read as a map of minimum handoff capability rather than a list of desirable topics.

\section{\large Skills map and role targets}

Titles do not usually break projects; handoffs do. For that reason, this section treats skills and roles through the artifacts that must move safely from one stage to the next. Table~\ref{tab:taxonomy} defines what \Lone, \Ltwo, and \Lthree\ mean in each domain. Table~\ref{tab:RoleDomainTargets} then assigns the minimum depth needed for common roles. The two tables are meant to be read together: one defines the level, and the other assigns the minimum level by role.

The abbreviations are defined once here to keep the tables readable. In Table~\ref{tab:RoleDomainTargets}, Th = Theory and Wave Optics; Sim = Simulation and V\&V; ID = Inverse Design; Fab = Fabrication; Test = Metrology and Test; Pkg = Packaging and Integration; Mfg = Manufacturing, Quality and Reliability; SW = Software and Reproducibility.

The most important threshold is usually \Ltwo. That is the point where a piece of work becomes handoff-ready. A field plot without convergence evidence, a layout without a layer map, or a measurement without reference normalization and uncertainty is still only part of the job. The tables below are intended to make that boundary visible.
\begin{table}[H]
\caption{Core domains and artifact-based descriptors. The table defines what \Lone, \Ltwo, and \Lthree\ mean in each domain.}
\label{tab:taxonomy}
\centering
\footnotesize
\renewcommand{\arraystretch}{1.10}
\begin{tabularx}{\textwidth}{>{\raggedright\arraybackslash}p{2.1cm} X X X}
\toprule
\textbf{Domain} & \Lone\ --- Guided & \Ltwo\ --- Independent & \Lthree\ --- Lead / Approve\\
\midrule
Theory \& Wave Optics &
Uses lens phase, grating relations, numerical aperture, and basic budget checks with guidance; states assumptions and units clearly. &
Translates system needs into device budgets; chooses scalar versus vector models; identifies dominant trade-offs and likely error sources. &
Sets modeling assumptions and review rules; defends budgets in technical reviews; mentors others on regime choice and error budgeting.\\
\multicolumn{4}{@{}l@{}}{\rule{0pt}{0.6em}}\\
Simulation, V\&V &
Runs reference models; records settings; performs basic convergence checks on provided cases. &
Chooses the solver, boundaries, and monitors; shows convergence and independent cross-checks; reports validity range and residual risk. &
Sets team verification standards; reviews model packages for release-driving decisions; decides when simulation evidence is sufficient.\\
\multicolumn{4}{@{}l@{}}{\rule{0pt}{0.6em}}\\
Inverse Design &
Runs an existing optimization flow; records seeds, settings, constraints, and outputs; interprets objective trends. &
Adds fabrication and robustness constraints; verifies the final design outside the optimizer; reports sensitivity to realistic bias and process windows. &
Owns optimization strategy, robustness policy, and traceability; decides when an optimized design is ready for handoff.\\
\multicolumn{4}{@{}l@{}}{\rule{0pt}{0.6em}}\\
Fabrication &
Reads travelers, layer maps, and DRC reports; understands lithography, etch, deposition, and safety basics. &
Prepares controlled release notes; defines process windows and control points; links geometry variation to optical change. &
Leads process integration and yield stabilization; approves process-ready releases, monitors, and deviation handling.\\
\multicolumn{4}{@{}l@{}}{\rule{0pt}{0.6em}}\\
Metrology \& Test &
Runs calibrated procedures; normalizes to references; computes key metrics; preserves raw data and metadata. &
Designs the test plan, uncertainty budget, and decision rule; correlates measured outputs to models and requirements. &
Owns qualification metrics and approves decision-ready reports; connects observed failures back to design, process, or packaging causes.\\
\multicolumn{4}{@{}l@{}}{\rule{0pt}{0.6em}}\\
Packaging \& Integration &
Understands alignment marks, fixtures, contamination control, and dominant sources of optical drift. &
Allocates alignment and drift budgets; plans environmental tests; links materials and assembly choices to optical performance. &
Owns the integration architecture and package-level acceptance tests; decides whether packaged hardware still meets optical intent.\\
\multicolumn{4}{@{}l@{}}{\rule{0pt}{0.6em}}\\
Manufacturing, Quality \& Reliability &
Uses revision control, change logs, and basic quality language correctly. &
Defines acceptance limits, reaction plans, and yield/excursion analysis; interprets reliability data in engineering terms. &
Owns release criteria, production readiness, deviation closure, and traceability across vendors or facilities.\\
\multicolumn{4}{@{}l@{}}{\rule{0pt}{0.6em}}\\
Software \& Reproducibility &
Uses version control and runnable notebooks or scripts; records environment and parameter settings. &
Builds reusable code with testing and data provenance; reproduces figures and tables from source data. &
Sets team coding and review standards; supports audits, reuse, and long-term maintainability.\\
\bottomrule
\end{tabularx}
\end{table}

\begin{table}[H]
\caption{Minimum domain depth for common roles. Read this table together with Table~\ref{tab:taxonomy}: Table~\ref{tab:taxonomy} defines the levels; this table assigns the minimum level by role.}
\label{tab:RoleDomainTargets}
\centering
\small
\setlength{\tabcolsep}{7.5pt}
\renewcommand{\arraystretch}{1.15}
\newcommand{\LoneC}{\cellcolor{p1}\Lone}
\newcommand{\LtwoC}{\cellcolor{p2}\Ltwo}
\newcommand{\LthreeC}{\cellcolor{p3}\Lthree}
\begin{tabular}{p{3.7cm} cccccccc}
\toprule
\textbf{Role} & \textbf{Th} & \textbf{Sim} & \textbf{ID} & \textbf{Fab} & \textbf{Test} & \textbf{Pkg} & \textbf{Mfg} & \textbf{SW}\\
\midrule
Meta/Diffractive Optics Design Engineer        & \LthreeC & \LtwoC   & \LtwoC   & \LtwoC   & \LtwoC   & \LtwoC   & \LoneC   & \LtwoC\\
Computational/Inverse Design Engineer          & \LtwoC   & \LthreeC & \LthreeC & \LtwoC   & \LoneC   & \LoneC   & \LoneC   & \LthreeC\\
Photonics Simulation Engineer                  & \LtwoC   & \LthreeC & \LoneC   & \LoneC   & \LtwoC   & \LoneC   & \LoneC   & \LtwoC\\
Process/Nanofab Engineer                       & \LoneC   & \LoneC   & \LoneC   & \LthreeC & \LtwoC   & \LtwoC   & \LtwoC   & \LoneC\\
Metrology/Test Engineer                        & \LtwoC   & \LtwoC   & \LoneC   & \LtwoC   & \LthreeC & \LtwoC   & \LtwoC   & \LtwoC\\
Packaging/Integration Engineer                 & \LoneC   & \LoneC   & \LoneC   & \LtwoC   & \LtwoC   & \LthreeC & \LtwoC   & \LoneC\\
Systems Engineer/Optical Architect             & \LthreeC & \LtwoC   & \LoneC   & \LtwoC   & \LtwoC   & \LtwoC   & \LtwoC   & \LtwoC\\
Manufacturing/Quality Engineer                 & \LoneC   & \LoneC   & \LoneC   & \LtwoC   & \LtwoC   & \LtwoC   & \LthreeC & \LtwoC\\
Applications/Product Engineer                  & \LtwoC   & \LoneC   & \LoneC   & \LoneC   & \LtwoC   & \LtwoC   & \LtwoC   & \LoneC\\
Technical Program Manager (TPM)                & \LoneC   & \LoneC   & \LoneC   & \LoneC   & \LoneC   & \LoneC   & \LtwoC   & \LtwoC\\
\bottomrule
\end{tabular}
\end{table}

Three practical takeaways are worth keeping in mind while reading these tables. First, \Ltwo\ is the threshold that usually determines whether a handoff is safe. Second, no role is completely local: design roles still need packaging and test literacy, while fabrication and metrology roles still need enough optics to understand optical impact. Third, programs with tight packaging, reliability, or high-volume manufacturing constraints should raise the packaging, metrology, and manufacturing targets above the minimum values shown here. Once the role targets are clear, the remaining question is educational: where do current programs undertrain these handoff-ready skills?

\section{\large Curriculum gaps and workforce development models}

Most optics and photonics programs teach wave physics better than they teach release discipline. The weak point is usually the handoff: turning a design into a release package, a test plan, and a decision-ready report. Four missing capabilities appear repeatedly: requirements traceability, model verification, uncertainty-based validation, and packaging-aware tolerancing. These gaps map directly to Gate~1, Gates~2--3, Gate~7, and Stage~8 in Fig.~\ref{fig:PdzWrkflw}.

Those gaps persist for structural reasons. They cut across courses, labs, and facilities. They often require cleanroom access, metrology time, or integration fixtures that are difficult to scale. They also work against common academic incentives, which tend to reward novelty or a single peak number more than repeatability, documentation, and controlled iteration.

 {Programs become easier to design when workforce development is organized as a small number of recurring models rather than as one catch-all curriculum. Table~\ref{tab:workforce_models} translates the workflow into three such pathways: a technician and manufacturing-test pathway, a design and foundry pathway, and a systems and productization pathway. These are not rigid tracks. They are practical ways to decide which artifacts a program must repeatedly teach and assess.}

\begin{table}[t]
\caption{ {Representative workforce-development models for flat optics. The table groups training by the kinds of handoffs and deliverables a learner must master, rather than by device label alone.}}
\label{tab:workforce_models}
\centering
\footnotesize
\renewcommand{\arraystretch}{1.08}
\setlength{\tabcolsep}{4pt}
\begin{tabularx}{\textwidth}{>{\raggedright\arraybackslash}p{2.9cm} >{\raggedright\arraybackslash}p{2.8cm} >{\raggedright\arraybackslash}X >{\raggedright\arraybackslash}p{3.5cm}}
\toprule
\textbf{Model} & \textbf{Typical audience} & \textbf{Strongest stages and domains} & \textbf{Signature deliverables}\\
\midrule
Technician / manufacturing / test & Two-year programs, upper-division lab sequences, early career manufacturing and test staff & Stages 5--8; fabrication, metrology, quality, reproducibility & Process travelers, calibration records, pass/fail reports, reaction plans, controlled data packages\\

Design / foundry / inverse design & Upper-division design courses, graduate students, fabless design teams & Stages 1--5; theory, simulation, inverse design, release discipline & Traceability matrix, verified model package, constraint sheet, layout release, DRC summary\\

Systems / productization / integration & Capstones, cross-functional professional training, product and applications teams & Stages 1--8; systems thinking, metrology, packaging, manufacturing handoff & Requirements budget, validation plan, stress-to-metric map, qualification report, release record\\
\bottomrule
\end{tabularx}
\end{table}

These models align well with existing efforts. AIM Photonics emphasizes design, testing, and packaging. JePPIX-style training emphasizes foundry-aware design and release discipline. NNCI and shared nanofabrication ecosystems emphasize safe process practice. ABET and CDIO reinforce the same educational logic: repeated design-implement experiences, aligned assessment, and evidence of engineering judgment \cite{AIMPhotonicsEWD,AIMPhotonicsSummerAcademy,JePPIX_PICDesignCourse,NNCI_NSF,ABET_EAC_2025_2026,CDIO_Standards_3_0,Biggs1996ConstructiveAlignment}.

The practical implication is not simply to add one more disconnected course. It is to build repeated artifact handoffs into existing courses, labs, and capstones so that students practice the same transfer they will later face in a company. The next section turns that idea into a concrete training blueprint.

\section{\large Training blueprint: educational translation of the workflow}

The classroom does not need to imitate a factory floor literally. It does need to teach the same habits of evidence, transfer, and review. Each module in Table~\ref{tab:Modules} ends with a simplified but authentic handoff artifact. The deliverable is lighter than an industry release package, but it tests the same habit: clear assumptions, reproducible analysis, and evidence that supports a decision.
\begin{table}[H]
\caption{Training blueprint organized as course modules. Each module defines a learning task, a student deliverable, and an assessment question. The table translates the workflow in Fig.~\ref{fig:PdzWrkflw} into teaching practice.}
\label{tab:Modules}
\centering
\footnotesize
\renewcommand{\arraystretch}{1.04}
\begin{tabularx}{\textwidth}{>{\raggedright\arraybackslash}p{2.9cm} X X >{\raggedright\arraybackslash}p{3.0cm}}
\toprule
\textbf{Module} & \textbf{What students practice} & \textbf{Student deliverables} & \textbf{Assessment question}\\
\midrule
A. Requirements and traceability &
Translate a system need into measurable device targets and test outputs. &
One-page requirement brief, traceability matrix, optical budget, assumptions log. &
Is the requirement clear and measurable?\\

B. Physical choice: DOE or metasurface &
Choose the right physical picture and state the model limits. &
Concept comparison memo, first design, parameter sweep, regime note. &
Is the chosen concept justified?\\

C. Model credibility &
Show that the model is stable, correctly normalized, and fit for decisions. &
Verified notebook, convergence study, energy check, one independent cross-check. &
Can another person trust the prediction?\\

D. Design under fabrication limits &
Encode minimum width, spacing, aspect ratio, and bias limits before release. &
Constraint sheet, reproducible optimization archive, robustness summary, verified candidate. &
Would the design survive realistic fabrication error?\\

E. Release discipline &
Turn the design into a package another team can use. &
Layout file, layer map, DRC summary, revision tag, release README. &
Could another team fabricate this without questions?\\

F. Validation and uncertainty &
Measure the right quantity, normalize it, estimate uncertainty, and make a decision. &
Test plan, raw dataset, calibration record, uncertainty budget, validation report. &
Does the report support pass/fail?\\

G. Packaging and drift &
Treat alignment and environment as optical constraints. &
Alignment budget, stress-to-metric map, qualification plan, final review package. &
Does the packaged part still meet the requirement?\\
\bottomrule
\end{tabularx}
\end{table}

This blueprint can support a capstone sequence, a laboratory sequence, a short-course series, or professional upskilling. The schedule can vary; the artifact logic should not. Students should be graded on clarity of assumptions, reproducibility of analysis, and strength of evidence, not only on the final performance number. That approach matches what recent roadmaps and metrology reviews identify as the real deployment bottlenecks: manufacturability, model credibility, calibrated validation, and packaging-aware integration \cite{Kuznetsov2024RoadmapACSP,Schulz2024RoadmapAPL,Yang2025CommercializingMetasurfaces,Kossowski2025MetrologyNPJ,Schubert2022StrictFoundry,Prince2004ActiveLearning,LavadoAnguera2024PBL}.

Section~8 defines what learners do. The next section addresses how universities and companies can adopt the framework and assess whether it is working.

\section{\large Implementation pathways and assessment}

Implementation is best treated as a system-design problem. First, define the target roles using Table~\ref{tab:RoleDomainTargets}. Second, identify which handoffs fail most often in the local environment using Fig.~\ref{fig:PdzWrkflw} and Table~\ref{tab:stagegates}. Third, select or bundle modules from Table~\ref{tab:Modules} so that students or new hires repeatedly practice those weak handoffs.

Universities and companies can use the same framework with different emphasis. Universities are likely to implement it as a capstone or two-course sequence with staged reviews: requirements review, model plan review, verification review, release review, test-readiness review, and final validation or qualification review. Companies are more likely to implement the same logic through onboarding and continuous training around real release gates. The artifact sequence remains the same; what changes is how close the exercises are to production hardware.

Assessment becomes clearer when every review uses the same small set of questions. Is the requirement clear? Is the model credible? Is the release reproducible? Do the data support the claim? Can the next team proceed safely? That rubric aligns naturally with ABET, CDIO, and constructive-alignment principles \cite{ABET_EAC_2025_2026,CDIO_Standards_3_0,Malmqvist2020CDIOUpdates,Biggs1996ConstructiveAlignment}. Program outcomes can then be tied to artifacts rather than impressions: time to first \Ltwo\ artifact, first-pass gate success rate, rerun success of computational submissions, the fraction of reports with complete uncertainty statements, and the number of late integration failures traced to missing upstream evidence.

With the implementation logic in place, the final section looks ahead at the trends most likely to push these requirements further upstream.

\section{\large Outlook: practical implementation guidance for universities and industry}

Flat optics has entered a deployment era. The limiting question is increasingly not whether a function can be demonstrated once, but whether it can be manufactured, measured, packaged, and qualified repeatedly with predictable yield and schedule. That shift changes what workforce readiness means. The most transferable skills are the ones that survive handoffs: measurable requirements, verified models, unambiguous releases, calibrated validation, and qualification plans that connect applied stress to optical metrics \cite{Kuznetsov2024RoadmapACSP,Schulz2024RoadmapAPL,Yang2025CommercializingMetasurfaces}.

Several near-term trends will keep pushing constraints upstream. Hybrid architectures that combine refractive optics, DOEs, and metasurfaces will become more common. Manufacturing-first design will continue to move earlier in the workflow. Decision-oriented metrology will matter more because release claims will be judged increasingly by uncertainty-aware evidence. Tunable platforms and AI-assisted design or measurement workflows only strengthen the need for a versioned digital thread that links requirements, simulation, release, fabrication data, and validation reports \cite{Zhang2020MetagratingCPD,Schubert2022StrictFoundry,Kossowski2025MetrologyNPJ,Lepeshov2021PhaseChange,Badloe2024AIEnhancedMetasurfaces}.

For emerging R1 and R2 programs, the most effective strategy is to build around artifacts rather than around an equipment wish list. A strong program does not need every fabrication and metrology tool in-house on day one. It needs a computation-first backbone that teaches verification and reproducibility, paired with a validation-first laboratory experience that teaches calibration, normalization, and decision rules. Shared cyberinfrastructure and teaching kits can make that model practical even when local infrastructure is limited \cite{ACCESS_CI,Thorlabs_PhysicsLab}.

Shared facilities can multiply that effort. Regional nanofabrication and photonics facilities, the NSF NNCI network, and DOE user facilities make it possible for universities without deep local infrastructure to teach real release, traveler, and validation practice \cite{CUNYASRC_PhotonicsFacility,CUNYASRC_Nanofabrication,NSF_NNCI,DOE_OS_UserFacilities,DOE_AccessModels,DOE_NSRCs,NSF_NQNI}. A useful model is to complete design, verification, and release artifacts locally, execute fabrication or measurement at a shared facility, and then complete the validation and pass-fail decision back on campus.

Industry benefits when workforce development is treated as upstream yield engineering. The highest-value contributions are specific and non-proprietary: clear requirement definitions, representative test fixtures or reference samples, example validation report templates, and capstone problems that end in a gate-ready package rather than in a single performance plot. Those contributions shorten onboarding and reduce late-stage iteration.

\section{\large Conclusion}

Meta-optics and diffractive optics now fail or succeed as much at the interfaces between stages as they do in the optical design itself. The most useful workforce language is therefore not a list of favorite tools or fashionable topics. It is a set of artifacts that can cross a handoff: a traceability matrix, a verified model package, a controlled release, a calibrated measurement report with uncertainty and a decision rule, an alignment or drift budget, and a qualification decision.

For companies, this framework supports clearer hiring and faster onboarding. For universities, it turns strong optics fundamentals into product-ready practice without rewriting an entire curriculum. For students and practicing engineers, it suggests a concrete portfolio: show one artifact for each major handoff, not just one final plot.  {The key question is no longer whether a flat optic can work once, but whether a team can hand off the evidence needed to make it work again.}

\section*{Disclosures}
The authors declare no conflicts of interest.

\section*{Code, Data, and Materials Availability}
No proprietary datasets are produced. The taxonomy, tables, and checklists in this paper are designed to be reused as templates in courses, onboarding, and internal documentation.

\section*{Acknowledgments}
The authors thank colleagues in meta-optics, diffractive optics, and engineering education for community resources and roadmaps that informed this tutorial.

\bibliographystyle{spiejour}
\bibliography{references}

\section*{Biographies}
\noindent Ingrid Torres is a postdoctoral researcher in electrical and computer engineering at Florida International University. Her work focuses on nanophotonics, metasurfaces, materials, and device fabrication, with current emphasis on cryogenic dielectric resonators, quantum photonics, and tunable metamaterial platforms for sensing and imaging.

\medskip
\noindent Alex Krasnok is an assistant professor of electrical and computer engineering at Florida International University, where he directs the Quantum Technology and Metamaterials Lab. His research spans quantum photonics, complex-frequency electrodynamics, metasurfaces, cryogenic and superconducting systems, and quantum sensing. He has authored more than 200 peer-reviewed publications and has led or contributed to federally funded projects in photonics, quantum technologies, and advanced electromagnetic systems.

\section*{Figure captions}
\noindent Figure 1. Historical development of diffractive optics and meta-optics (1948--2026). Lane A tracks holography and diffractive optics toward mature DOE-based industrial components. Lane B tracks metamaterials and metasurfaces toward metalenses and multifunctional meta-systems. The shaded deployment era highlights the shift from demonstrations to scalable fabrication, manufacturable design, calibrated metrology with uncertainty, packaging, reliability, and commercialization.

\medskip
\noindent Figure 2. Meta-optics and diffractive optics pursue the same engineering goal---shaping the optical field across a surface---but do so with different physical building blocks. Metasurfaces rely on wavelength-scale scatterers and naturally bring amplitude and polarization into the design space. Diffractive optics relies on engineered phase or amplitude profiles and often begins with Fourier-optics design intuition. In products, the two overlap because both must eventually satisfy the same constraints on fabrication, metrology, packaging, yield, cost, and schedule.

\medskip
\noindent Figure 3. High-level stage-gate workflow for meta-optics and diffractive optics. The figure shows the stage order, the redesign loop, and the position of Gates~1--8. Gate~7 sits between metrology/validation and packaging/qualification/release. Detailed artifacts and exit questions are listed in Table~\ref{tab:stagegates}.

\medskip
\noindent Figure 4. Geometry used in Eqs.~(\ref{eq:fresnel_number})--(\ref{eq:phase_gradient_angle}). The flat optic lies in the \(x\)-\(y\) plane, the surface normal is \(+\hat{z}\), \(z\) is the propagation distance measured along that normal, \(a\) is the half-aperture used in the Fresnel number, and \(\theta_{\mathrm{in}}\) and \(\theta_{\mathrm{out}}\) are measured from the surface normal. The schematic defines a generic surface-based angle convention; the same notation supports transmission or reflection once the corresponding sign convention is chosen.

\end{document}